\documentclass[a4paper,10pt,5p,preprint]{elsarticle}

\usepackage[table,xcdraw]{xcolor}
\usepackage{graphicx}
\usepackage[caption=false]{subfig}
\usepackage{amsmath}
\usepackage{amsfonts}
\usepackage{url}
\usepackage{multirow}
\usepackage{array}
\usepackage{bbold}
\usepackage{booktabs}
\usepackage{breakcites}
\usepackage{epstopdf}
\usepackage{morefloats}
\usepackage[pdfpagelabels, hypertexnames=true, plainpages=false, naturalnames=false, breaklinks=true]{hyperref}

\newcolumntype{C}[1]{>{\centering\arraybackslash}m{#1}}

\newcommand{\change}[1]{#1}

\journal{Environmental Modelling \& Software}

\biboptions{authoryear}

\begin{document}

\begin{frontmatter}

\title{A copula-based sensitivity analysis method and its application to a North Sea sediment transport model}

\author[TuDelftCiTG]{Matei \c{T}ene\corref{correspondingauthor}}
\cortext[correspondingauthor]{Corresponding author}
\ead{m.tene@tudelft.nl}

\author[Deltares]{Dana E. Stuparu}
\ead{dana.stuparu@deltares.nl}

\author[TuDelftEWI]{Dorota Kurowicka}
\ead{d.kurowicka@tudelft.nl}

\author[Deltares]{Ghada Y. El Serafy}
\ead{ghada.elserafy@deltares.nl}

\address[TuDelftCiTG]{Department of Geoscience and Engineering, Delft University of Technology\\P.O. Box 5048, 2600 GA Delft, The Netherlands.}
\address[Deltares]{Deltares, P.O. Box 177, 2600 MH Delft, The Netherlands}
\address[TuDelftEWI]{Department of Applied Mathematics, Delft University of Technology\\P.O. Box 5031, 2600 GA Delft, The Netherlands.}

\begin{abstract}
This paper describes a novel sensitivity analysis method, able to handle dependency relationships between model parameters. The starting point is the popular \cite{morris} algorithm, which was initially devised under the assumption of parameter independence. This important limitation is tackled by allowing the user to incorporate dependency information through a copula. The set of model runs obtained using latin hypercube sampling, are then used for deriving appropriate sensitivity measures.

\noindent Delft3D-WAQ \citep{delft3d} is a sediment transport model with strong correlations between input parameters. Despite this, the parameter ranking obtained with the newly proposed method is in accordance with the knowledge obtained from expert judgment. However, under the same conditions, the classic Morris method elicits its results from model runs which break the assumptions of the underlying physical processes. This leads to the conclusion that the proposed extension is superior to the classic Morris algorithm and can accommodate a wide range of use cases.
\end{abstract}


\end{frontmatter}


\section{Introduction}
\label{sec:intro}

Suspended particulate matter (SPM) is composed of fine-grained particles of both inorganic and organic origin, which are suspended in the water column. This material plays an important role in the ecology of coastal areas, as it influences the underwater light conditions (directly connected to the phytoplankton growth), the amount of nutrients in the water, the material transfers to the seabed and other environmental processes. As such, the SPM concentration plays a crucial role in the dynamics of aquatic ecosystems. At the same time, the increasing number of human activities along the shorelines (fishing, sand and gravel extraction, tourism, industry) often disturb the natural equilibrium of the sediment transport processes. To assess and monitor the possible impacts on the sediment transport patterns, models are used to estimate and forecast their movement, under the combined action of both natural factors and human interference.

The current study concerns the southern North Sea area, a marine system significantly affected by SPM, since it receives the run-off from major rivers and coastal industries. This area has seen rising interest in the scientific community \citep{fettweis2006,pietrzak2011}, which has led to the continuous development of the Delft3D-WAQ \citep{delft3d} sediment transport and water quality model \citep[see][]{elSerafy,blaas2007}. Delft3D-WAQ makes use of the hydrodynamic conditions (velocities, discharges, water levels, vertical eddy viscosity and vertical eddy diffusivity) and wave characteristics (important in the sediment re-suspension and settling) to simulate the complex interplay between the hydrodynamic, chemical and biological processes involved in the sediment transport system.

However, calibrating this model is made difficult by the large number of input parameters, some of which are strongly correlated, due to physical constraints. Also, the high running time for one simulation  - approximately 11 hours in full resolution and 3 hours on a coarse grid - imposes additional restrictions on the calibration efforts. This gave rise to the question of whether the model parameters can be ranked, such that the calibration process can be focused on only the subset \change{to which the output is most sensitive}. The remaining parameters can be fixed to their maximum likelihood values (determined, for example, using an expert judgment exercise).

\change{According to \cite{vanGriensven2006}, o\-ver-pa\-ra\-me\-ter\-i\-za\-tion is a widespread problem for environmental models. At the same time, \cite{shin2013} point out that only few studies in the literature \citep[see, e.g.,][]{schmid2003,francos2003,shen2008,plecha2010,kurniawan2011} employ sensitivity analysis methods \citep{saltelli2000,makler} to rank parameters and identify redundancies. Among these, the method developed by \cite{morris} is especially popular \citep{campolongo1997b,portilla,arabi} due to its simplicity and computational efficiency.} However, in its initial formulation, the method assumes independence between model parameters. This can be a limiting factor, since, in many cases, the physically-induced dependencies can not be overlooked. \change{For example, in \cite{campolongo1997a} the authors had to eliminate certain parameters from their analysis, specifically because of this limitation. Also, in \citep{Salacinskaetal(2009)}, the sensitivity of the simulated chlorophyll-a concentration to a subset of ecologically significant input factors has been carried out with the use of the Morris method and later enriched by the computation of the correlation ratios of the selected parameters on the model response at a few selected locations in the domain. The second step was crucial to obtain results in agreement with expert knowledge of the ecological processes in the North Sea.}

\change{This paper proposes an extension to Morris' method which opens the possibility to control the sampling pattern of the necessary model runs for sensitivity analysis, based on prior information about dependencies between model parameters. More specifically, this work incorporates this information into the sampling strategy of the elementary effects in the Morris method. The dependencies can be specified in terms of parameter correlations or, more precisely, by providing their joint distribution. This leads to the construction of the corresponding copula \cite{nelsen2007} -- a joint distribution on the unit hypercube with uniform marginals -- which is, finally, used to determine the set of model simulations required to conduct the sensitive analysis study.}

The application on the computationally expensive Delft 3D-WAQ sediment transport model confirms that the method is able to provide physically sound results regarding the parameter ranking, even in cases where the feasible number of simulations is limited. This confirms the relevance of the method in identifying the parameters having the strongest impact on the variability of the model predictions.

The content of the paper is structured as follows. First, the Delft3D-WAQ sediment transport model and the dependence relationships between the governing model parameters are introduced. Next, the classic Morris sensitivity analysis method is reviewed, which presents the opportunity to devise a geometrical reinterpretation of its elementary effect sampling strategy, separating it into three successive stages. This new insight leads to the formulation of mechanisms to constrain each stage of Morris method, by incorporating the prior information about parameter dependencies in the form of a joint distribution with corresponding copula. In Section~\ref{sec:copula}, the basic theory concerning copulas is summarized and Section~\ref{sec:extended_morris} presents the newly developed copula-based Morris method. Finally, both methods are applied to the Delft3D-WAQ model and the results are compared, leading to the conclusions.


\section{The Delft3D-WAQ sediment transport model for the southern North Sea}
\label{sec:delft3d}

With an extensive history of maritime commerce, the North Sea is one of the most intensively traversed sea areas. It is bordered by highly industrialized and densely populated countries, which are actively engaged in mineral extraction, diking, land reclamation and other activities. The main sources of sediments are the Dover straits, the Atlantic Ocean, river bed and coastal erosion \citep{kamel2013}. The SPM concentration varies in both time and space, as a response of the seabed to the hydro-meteorological forces that result from the interaction between river inflows, waves, winds, currents and external factors.

For example, the breaking waves in the near-shore areas, together with various horizontal and vertical current patterns are constantly transporting beach sediments. Sometimes, this transport results in only a local rearrangement of sand. However, under certain conditions, extensive displacements of sediments along the shore take place, possibly moving hundreds of thousands of cubic meters of sand along the coast each year.  During calm weather conditions, the SPM settles and mixes with the upper bed layers. Subsequently, strong near-bed currents, generated by tides or high surface waves, can trigger the resuspension of the SPM from the seabed into the water column.

\begin{figure}%
\centering%
\includegraphics[width=\columnwidth]{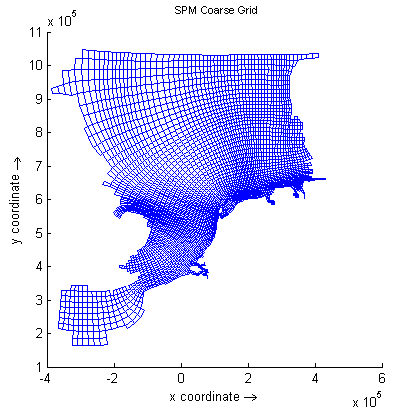}%
\caption{Delft3D-WAQ spatial discretization grid for the North Sea.}%
\label{fig:grid}%
\end{figure}

The Delft3D-WAQ \citep{delft3d} model is capable to describe the erosion, transport and deposition of SPM in the southern North Sea with a good degree of accuracy \citep{elSerafy}. In the model, SPM consists of three different fractions \citep{jimenez}: medium ($\text{IM}_1$, diameter 40 $\mu m$), coarse ($\text{IM}_2$, diameter 15 $\mu m$) and fine sediments ($\text{IM}_3$, diameter 1 $\mu m$). \change{These appellations are a Delft3D internal name and will be used to refer to three sediment types throughout the remainder of this paper.}  The model computes the convection-diffusion, settling and resuspension of the three silt fractions of SPM, given the transport velocities, mixing coefficients and bed shear stress adopted from the hydrodynamic and wave models. The spatial domain is covered by an orthogonal grid of $134 \times 165$ cells, with a resolution that varies between $2 \times 2 \  km^2$ in the coastal zone and $20 \times 20 \  km^2$ further offshore, as illustrated in Fig.~\ref{fig:grid}. Also, in order to capture the vertical structure of the flow, together with the stratification and mixing of SPM caused by the tidal influence in the domain, the water depth is modeled by 12 \change{so-called} \emph{sigma layers}, with different thicknesses (increased resolution near the seabed). The water surface is represented by the first layer and represents $4\%$ of the column depth.

Recently, Delft3D-WAQ has been extended with an improved parametrization of the resuspension and buffering of the silt fractions (related to both $\text{IM}_2$ and $\text{IM}_1$) from the seabed \citep{vanKessel}. This parametrization enables a more realistic description of the periodic and relatively limited resuspension during the tidal cycle and the massive resuspension from deeper bed layers observed during high wave events \citep{elSerafy}. Only the main features of this approach are described below; more details can be found in \cite{vanKessel}.

\begin{figure}%
\change{\centering%
\includegraphics[width=\columnwidth]{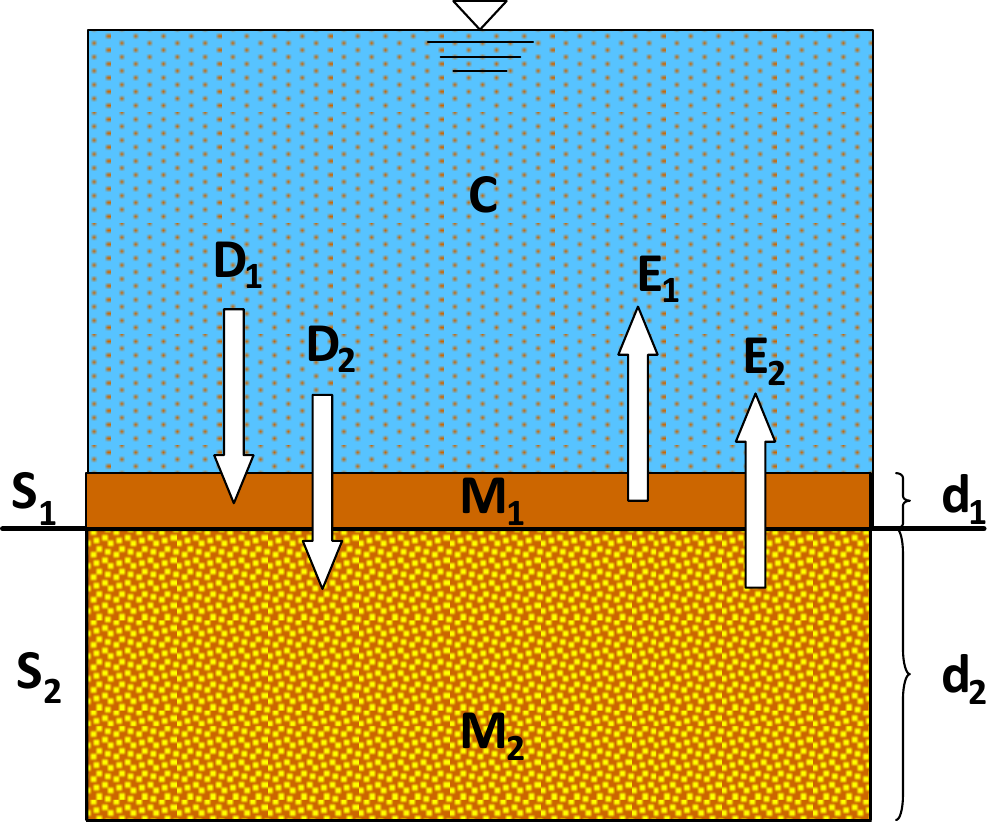}%
\caption{Schematic representation of the Delft3D-WAQ buffer model. Layer $S_1$ is the thin fluffy mud layer of thickness $d_1$, while the layer $S_2$ is the sandy sea bed infiltrated with fines of thickness $d_2$. $D_j$ is the deposition flux towards layer $S_j$, $E_j$ is the erosion flux from layer $S_j$, $M_j$ is the mass sediment fraction in ($j \in \{1,2\}$) and $C$ is the SPM concentration.}}%
\label{fig:buffer}%
\end{figure}

The buffer model contains two bed layers, each interacting with the water column in a specific way (Fig.~\ref{fig:buffer}). The first layer, denoted as $S_1$, is a thin fluffy mud layer that is easily resuspended by tidal currents. On the other hand, the sandy buffer layer, $S_2$, can store fines for a longer time and releases SPM only during highly dynamic conditions, such as spring tides or storms. Both layers interact with the water column, but with different rates, depending on the different physical processes involved in either settling or resuspension mechanisms.

The deposition towards the layers $S_1$ and $S_2$ is influenced by the settling velocity \change{$V_{\text{sed IM}_i}~ [m/day]$} and the saturation factor \change{$Fr_{\text{IM}_i\text{ sed S}_2}$}, which distributes the flux to the seabed. The main equations describing this process are:
\begin{align}%
& D_{1\text{ IM}_i} = (1 - Fr_{\text{IM}_i\text{ sed S}_2})~ V_{\text{sed IM}_i} C_{\text{IM}_i} \\%
& D_{2\text{ IM}_i} = Fr_{\text{IM}_i\text{ sed S}_2}~ V_{\text{sed IM}_i} C_{\text{IM}_i},%
\end{align}%
where $C_{\text{IM}_i}$ is the concentration of the inorganic fraction, $\text{IM}_i$ ($i \in \{1,2,3\}$).

Under certain conditions, resuspension events from the two layers occur. For the fluffy mud layer $S_1$, the resuspension of the SPM fractions is proportional to the respective critical shear stress levels \change{$\tau_{\text{cr S}_1\text{ IM}_i}~ [Pa]$}, as well as the resuspension rates \change{$V_{\text{res IM}_i}~ [1/day]$,
\begin{equation}%
E_{1\text{ IM}_i} = V_{\text{res IM}_i}~ M_{1\text{ IM}_i} \left( \frac{\tau}{\tau_{\text{cr S}_1\text{ IM}_i}} - 1 \right),%
\end{equation}%
where $\tau~ [Pa] $ is the bottom shear stress and $M_{1\text{ IM}_i}~ [g / m^2]$ is the mass sediment fraction.} For the buffer layer $S_2$, the fines can be mobilized only beyond critical mobilization conditions. Thus, the erosion process is mostly influenced by the critical shear stress \change{$\tau_\text{Shields}~ [Pa]$} and the overall pick up factor \change{$Fact_\text{res Pup} [kg/m^2/s]$,
\begin{equation}%
E_{1\text{ IM}_2} = Fact_\text{res Pup}~ M_{2\text{ IM}_i} \left( \frac{\tau}{\tau_\text{Shields}} - 1 \right)^{1.5},%
\end{equation}%
where $M_{2\text{ IM}_i}~ [g / m^2]$ is the layer's mass sediment fraction and the exponent is due to the empirical pick-up function for a sandy seabed from \cite{VanRijn}.}

These parameters and their relationships are further detailed in the next paragraph. \change{The results presented in this paper were obtained using version 4.5208 (released on 12-08-2010) of the Delft3D-WAQ model.}


\subsection{Parameters and dependencies}

\change{The Delft3D-WAQ model captures the effect of complex mechanical, chemical and biological processes involved in the sediment transport system. Consequently, a total of 71 input parameters need to be specified in order to setup a simulation scenario. However, for the purposes of the current study, only the 14 parameters which govern the deposition/resuspension processes described earlier, will be considered for sensitivity analysis. They are listed in Table~\ref{tab:params}, along with their expert elicited value ranges see \citep{vanKessel}. The baseline values represent the model parametrization before the present study. The feasible ranges of the parameters were estimated based on measurements, such as the average setting velocity of mud, critical shear stress for erosion and erosion rate parameter under given conditions.  For further information see \citep{WinterwerpKesteren(2004),Gryeretal(2006),Fettweis(2008)}. }

\begin{table*}%
\centering%
\change{\caption{Delft3D-WAQ deposition and erosion parameters, with ranges and baseline values.\label{tab:params}}}%
\begin{tabular}{ C{0.14\textwidth} C{0.13\textwidth} C{0.13\textwidth} C{0.13\textwidth} C{0.12\textwidth} C{0.19\textwidth} } \midrule%
\bf{Parameter} & \bf{Minimum} & \bf{Baseline} & \bf{Maximum} & \change{\bf{Unit}} & \change{\bf{Description}} \\ \midrule%
\change{$V_{\text{sed IM}_1}$}				& 5.04		& 10.8		& 43.2		& \multirow{3}{\linewidth}{\centering\change{$m/day$}	}	& \multirow{3}{\linewidth}{\centering\change{Deposition velocities}} \\ \cmidrule{1-4}%
\change{$V_{\text{sed IM}_2}$}				& 43.2		& 86.4		& 172.8		& & \\ \cmidrule{1-4}%
\change{$V_{\text{sed IM}_3}$}				& 0.1		& 0.1		& 5.04		& & \\ \midrule%
\change{$Fr_{\text{IM}_1\text{ sed S}_2}$}	& 0.05		& 0.15		& 0.4		& \multirow{3}{\linewidth}{\centering\change{$-$}}			& \multirow{3}{\linewidth}{\centering\change{$S_2$ deposition fractions}} \\ \cmidrule{1-4}%
\change{$Fr_{\text{IM}_2\text{ sed S}_2}$}	& 0.05		& 0.15		& 0.4		& & \\ \cmidrule{1-4}%
\change{$Fr_{\text{IM}_3\text{ sed S}_2}$}	& 0.05		& 0.15		& 0.4		& & \\ \midrule%
\change{$V_{\text{res IM}_1}$}				& 0.05		& 0.2		& 0.5		& \multirow{3}{\linewidth}{\centering\change{$1/day$}}		& \multirow{3}{\linewidth}{\centering\change{$S_1$ erosion velocities}} \\ \cmidrule{1-4}%
\change{$V_{\text{res IM}_2}$}				& 0.2		& 1			& 1.2		& & \\ \cmidrule{1-4}%
\change{$V_{\text{res IM}_3}$}				& 0.2		& 1			& 1.2		& & \\ \midrule%
\change{$Fact_\text{res Pup}$}				& 8e-9		& 3e-8		& 8e-8		& \change{$kg/m^2/s$}									& \change{$S_2$ erosion rate} \\ \midrule%
\change{$\tau_{\text{cr S}_1\text{ IM}_1}$}	& 0.05		& 0.1		& 0.2		& \multirow{3}{\linewidth}{\centering\change{$Pa$}}		& \multirow{3}{\linewidth}{\centering\change{$S_1$ erosion crit. shear stresses}} \\ \cmidrule{1-4}%
\change{$\tau_{\text{cr S}_1\text{ IM}_2}$}	& 0.05		& 0.1		& 0.2		& & \\ \cmidrule{1-4}%
\change{$\tau_{\text{cr S}_1\text{ IM}_3}$}	& 0.05		& 0.1		& 0.2		& & \\ \midrule%
\change{$\tau_\text{Shields}$}				& 0.4		& 0.8		& 1.2		& \change{$Pa$}										& \change{$S_2$ erosion crit. shear stress}\\ \midrule%
\end{tabular}%
\end{table*}%

The values of the 14 parameters need to respect the physical laws and empirical relations governing the fluxes of sediment within and between the water column and the seabed. More specifically, the long term equilibrium between the buffer capacity (sediment in the $S_2$ layer) and the water column needs to be preserved. Otherwise, the model would result in unrealistic outputs, for example localized accumulation or disappearance of sediments to/from the seabed. This necessity has resulted in a dependence structure between the model parameters, further described by the following relationships:

\begin{itemize}%
	\item an increase in parameter $V_{\text{sed IM}_i}$, needs to be accompanied by a decrease in parameter $Fr_{\text{IM}_i\text{ sed S}_2}$ (or vice-versa), so that the settling into layer $S_2$ is roughly preserved and the annual equilibrium is respected for each fraction $i \in \{1,2,3\}$;%
	\item the parameters $\tau_{\text{cr S}_1\text{ IM}_i}$ and $V_{\text{res IM}_i}$ need to increase or decrease simultaneously, such that the year-average resuspension from layer $S_1$ is roughly conserved for each fraction $i \in \{1,2,3\}$;%
	\item parameters $\tau_\text{Shields}$ and $Fact_\text{res Pup}$ need to increase or decrease simultaneously, so that that the year-average resuspension from layer $S_2$ remains equal.%
\end{itemize}%

This leads to the specification of 7 pairs (as given in Table~\ref{tab:paramsCor}), each pair formed by two parameters which:
\begin{enumerate}%
\item \label{psetup1} are completely rank-correlated%
\item \label{psetup2} vary in the same or opposite directions (according to the rank-correlation)%
\item \label{psetup3} vary simultaneously%
\end{enumerate}%
\change{Correlations between parameters belonging to different pairs are considered by experts as insignificant, hence independence is assumed. This amounts to a very sparse correlation matrix, with 7 non-zero correlations.}

\begin{table}[htp!]%
\centering%
\change{\caption{Completely rank-correlated pairs of parameters.\label{tab:paramsCor}}}%
\begin{tabular}{ c c } \midrule%
\bf{Pair} & \bf{Rank correlation} \\ \midrule%
\change{$V_{\text{sed IM}_1}$ -- $Fr_{\text{IM}_1\text{ sed S}_2}$} & -1 \\%
\change{$V_{\text{sed IM}_2}$ -- $Fr_{\text{IM}_2\text{ sed S}_2}$} & -1 \\%
\change{$V_{\text{sed IM}_3}$ -- $Fr_{\text{IM}_3\text{ sed S}_2}$} & -1 \\ %
\change{$V_{\text{res IM}_1}$ -- $\tau_{\text{cr S}_1\text{ IM}_1}$} & 1 \\%
\change{$V_{\text{res IM}_2}$ -- $\tau_{\text{cr S}_1\text{ IM}_2}$} & 1 \\%
\change{$V_{\text{res IM}_3}$ -- $\tau_{\text{cr S}_1\text{ IM}_3}$} & 1 \\%
\change{$\tau_\text{Shields}$ -- $Fact_\text{res Pup}$} & 1 \\ \midrule%
\end{tabular}%
\end{table}%


\subsection{Model output and MERIS Remote Sensing SPM}

The purpose of the sensitivity analysis is to identify the most important \change{deposition/erosion} parameters to be later used to calibrate the model against measured data. For this purpose, this paragraph will introduce a suitable sensitivity objective function.

\begin{figure*}%
\centering%
\includegraphics[width=0.75\textwidth]{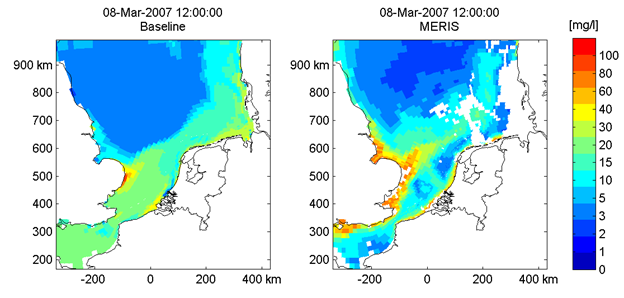}%
\caption{Comparison between the MERIS data and the corresponding Delft3D-WAQ simulation result.}%
\label{fig:ModelMeris}%
\end{figure*}

The model computes the total SPM concentration in each water surface grid cell on an hourly basis (calculated as the summation of the concentration of the three sediment fractions). In addition to this, SPM measurements retrieved from the optical remote sensing system ESA MERIS are available. This system supplies data from the visible, upper part of the water column, during the overpass of the Envisat satellite over the North Sea, occurring nominally once per day between 9:00 and 12:00 AM UTC. As SPM is a natural constituent of water, it affects the color of the sea. Therefore, the SPM concentrations in the water surface layer (several meters) can be derived from satellite snapshots, using the VU-IVM HYDROPT algorithm \citep{eleveld2008}. However, some SPM pixels need to be rejected for technical or quality reasons (cloudiness, land, unreliable retrieval, etc.) and have, thus, been removed from the measurements data set \citep{eleveld2008}. Fig.~\ref{fig:ModelMeris} illustrates an example of the MERIS data versus the model simulation results for the surface layer at the same time instance.

This allows the model error, $\epsilon$, to be defined as the spatial and temporal mean of the absolute differences between the model prediction and the MERIS data. If a measurement is not available in a given grid cell and time instance, that specific model output is discarded from the computation. Mathematically, this reads:
\begin{equation}%
\epsilon = \frac{1}{N} \sum_{i=1}^N | \text{Model}_{i} - \text{MERIS}_{i} |%
\end{equation}%
where $N$ is the number of measurements (in both time and space). \change{ Note that the sampling by the MERIS sensor is irregular in time and space, mainly due to factors such as the low sun angle, cloudy weather and rough sea states.  Therefore, the number of available measurements varies in time and space, with an average percentage coverage of the model domain of up to 60\%. The estimated concentrations range from 0 to 100 mg/l.  Using this function as sensitivity measure allows an assessment of the impact of each parameter on the ability of the model to forecast SPM concentrations.}

The results of the sensitivity study will be detailed in the following paragraphs.

\section{The classic Morris method}
\label{sec:classic_morris}

In this section, the concept of the classic Morris method is briefly presented, followed by a detailed discussion on the interpretation of the Morris sensitivity measures.

Given a model, $M$, with $n$ model parameters, $x = [x_1, \dots, x_n]$, the goal of the Morris method is to rank the model parameters according to their average effect on a particular model output. The method explores all model parameters, with a so called \textit{one-at-a-time} (OAT) design. More precisely, the model parameters are varied in turn and the effect each variation has on the output is then measured. This is done using the so called \emph{elementary effects}, which quantify the variation of the model output due to the variation in the model parameters.

This technique enables the identification of the model parameters $x_j$ affecting the output in a way that is: (a) negligible, (b) linear and additive, (c) nonlinear or involved in interactions with other parameters \citep{campolongo2007}. Note that in the case that the model has $m>1$ outputs, $y_1, y_2, \dots, y_m$, then, according to \cite{shan}, the effects can either be measured separately for each $y_k$ (the \emph{split method}) or in terms of a scalar-valued function of the $y_k$, \change{also referred to as \emph{quantity of interest}} (for example, an average or a norm).

After performing the sensitivity analysis, efforts can then be focused on calibration and fine-tuning of the parameters in category (c), while keeping the other parameters fixed to predefined values. Therefore, in its classic formulation, the Morris method is, essentially, a screening technique.


\subsection{Elementary effect analysis}
\label{sec:elem_effect}

\begin{figure}%
\centering%
\includegraphics[width=\columnwidth]{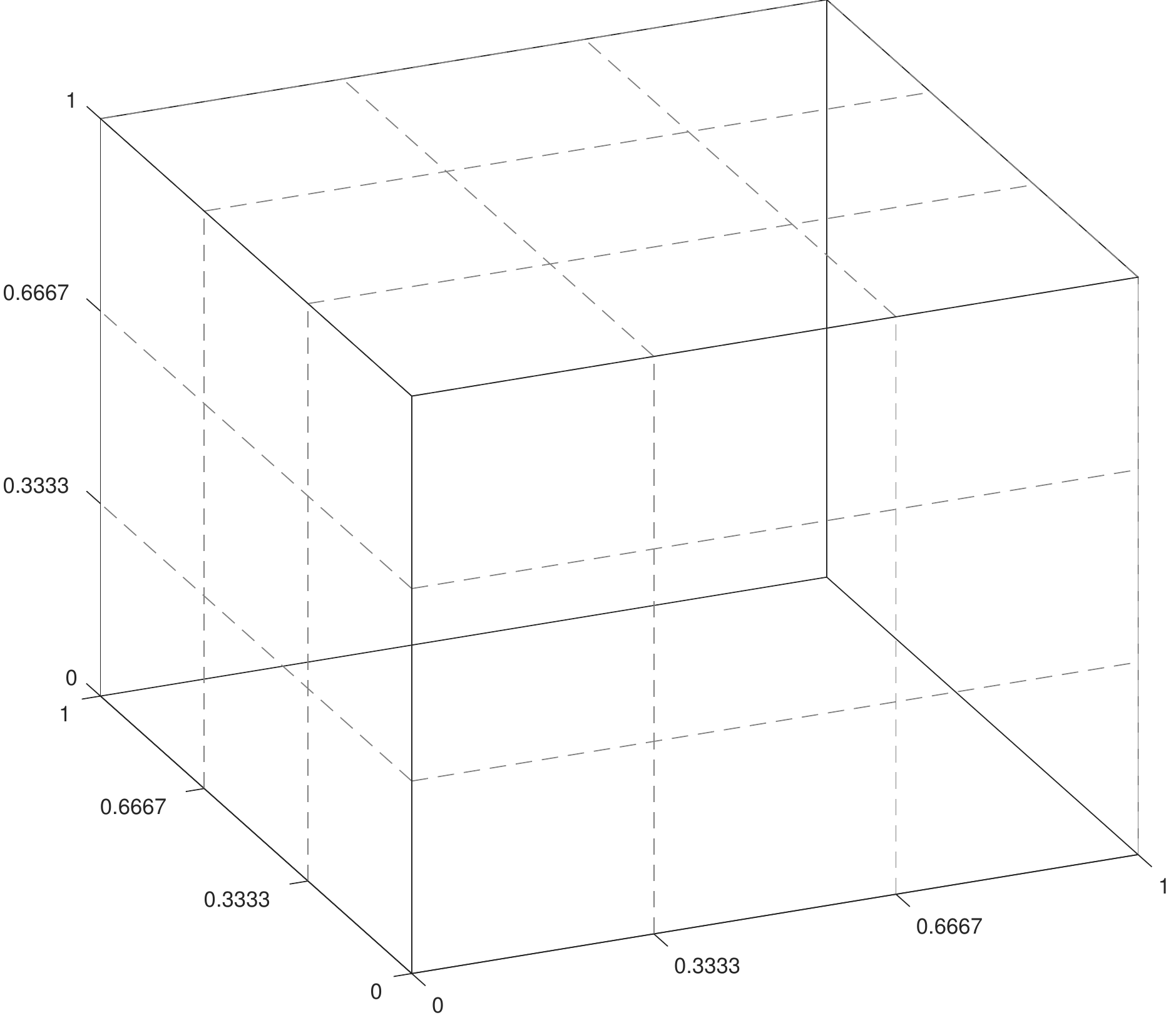}%
\caption{Unit hypercube representation of the parameter space for $n=3$  parameters and $p=4$ discretization levels.}%
\label{fig:hypercube}%
\end{figure}

The Morris method \citep{morris} determines the statistics of the, so-called, \emph{elementary effects} $d_j$, defined as
\begin{equation}%
d_j = \frac{M(x_1,\dots,x_{j-1},x_j + \delta,x_{j+1},\dots,x_n) - M(x_1,\dots,x_n)}{\delta}%
\label{eq:elem_effect}%
\end{equation}%
which serves as an approximation of the partial derivative of $M$ with respect to $x_j$. In order to evaluate $d_j$ independently of the parameter ranges, each $x_j$ is first scaled to $[0,1]$. This maps the parameter space to a unit hypercube, $[0,1]^n$, which is subsequently discretized in $p$ levels (an example is illustrated in Fig.~\ref{fig:hypercube}). The Morris step,
\begin{equation}%
\delta = \frac{s}{p-1} \hspace{2cm} s \in \{1, \dots, p-1\}%
\label{eq:step}%
\end{equation}%
represents the magnitude of the variation and is chosen as a multiple of the grid cell size, $\frac{1}{p-1}$.

In order to measure the \textit{average effect} of the parameter variation on the model output, elementary effects are calculated $r$ times for each parameter at randomly chosen positions on the grid. This allows for the computation of two sensitivity measures, the elementary mean and standard deviation:
\begin{equation}%
\mu_j = \frac{1}{r} \sum_{i=1}^r {d_j}^{(i)} \hspace{1cm} \sigma_j = \sqrt{\frac{1}{r-1} \sum_{i=1}^r \left( {d_j}^{(i)} - \mu_j \right)^2}%
\label{eq:measure1}%
\end{equation}%
which provide insight into the relative \change{sensitivity to} $x_j$. \\
Other sensitivity measures could be defined, for example, \cite{portilla} use the value of $\sqrt{{\mu_j}^2 + {\sigma_j}^2}$ to build a ranking of model parameters, while \cite{campolongo2007} recommend using the absolute elementary mean,
\begin{equation}%
\mu_j^* = \frac{1}{r} \sum_{i=1}^r \left| {d_j}^{(i)} \right| \hspace{2cm} j=1,\dots,n%
\label{eq:measure2}%
\end{equation}%
instead of $\mu_j$, in order to better capture elementary effects of opposing sign (which cancel each other out in the calculation of $\mu$).

The interpretation of $\mu_j$, $\mu_j^*$ and $\sigma_j$ in assessing the overall influence of parameter $x_j$ on the model output, is as follows. If $\mu_j$ has a high amplitude, it implies not only that the parameter has a large effect on the output, but also that the sign of this effect does not vary significantly over model simulations. Meanwhile, in the case that $\mu_j$ is relatively low and $\mu_j^*$ is high, $x_j$ has effects are of opposing sign, varying with the point of evaluation. In addition, if $\sigma_j$ is high, then the elementary effects relative to this parameter are significantly different from each other. This means that the value of $x_j$'s elementary effects are strongly dependent upon the choice of the point in the input space where it is evaluated, i.e., by the choice of the other parameters’ values. One may, therefore, conclude that this parameter has a high interaction with other parameters. On the other hand, a low value of $\sigma_j$ indicates nearly constant values of the elementary effects, therefore implying that the model is almost linearly dependent on $x_j$.

\begin{figure}%
\centering%
\subfloat[randomly sampled elementary effects]{\includegraphics[width=\columnwidth]{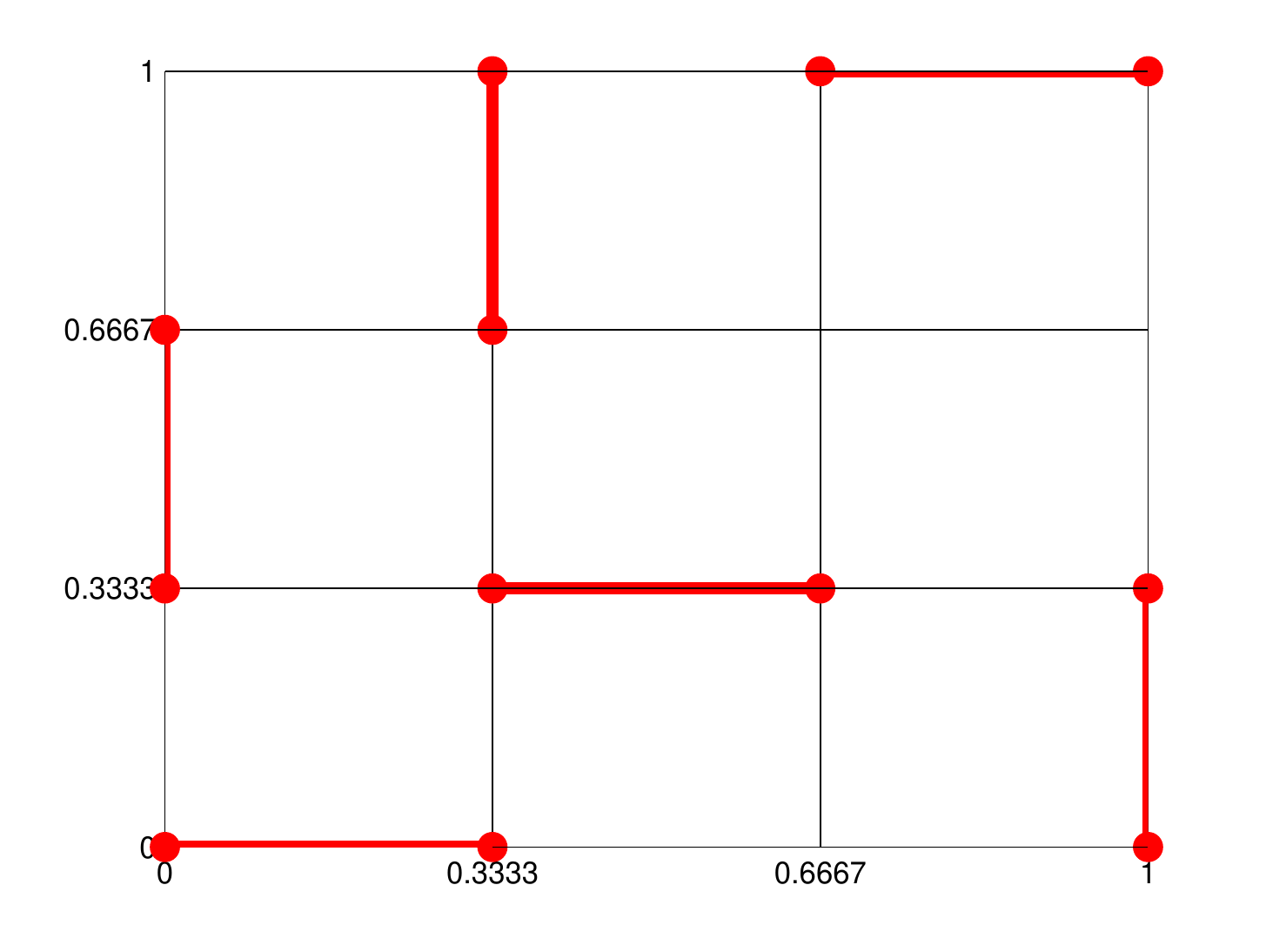}\label{fig:elem_effects}}\\%
\subfloat[effects grouped in elementary paths]{\includegraphics[width=\columnwidth]{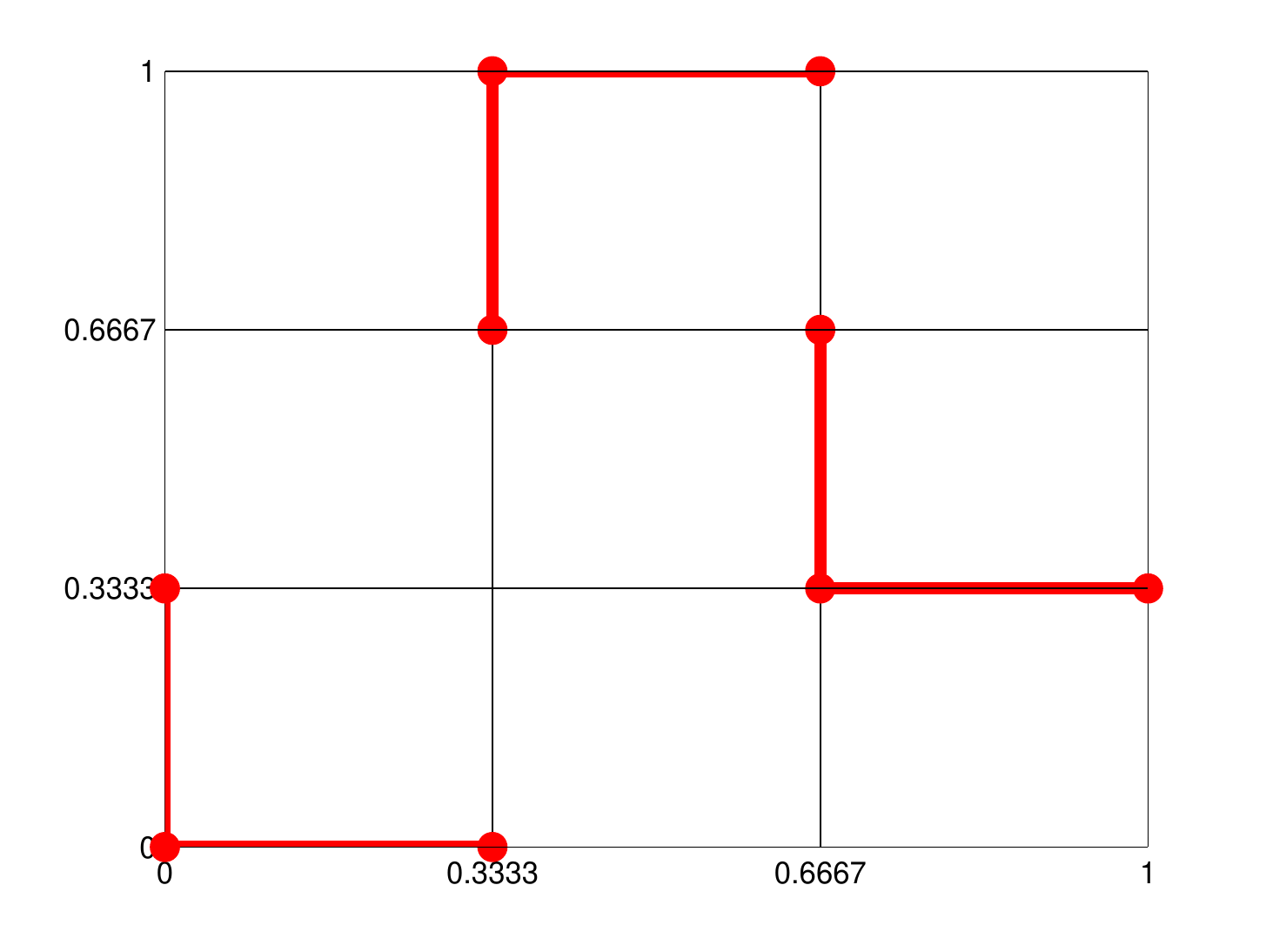}\label{fig:elem_paths}}%
\caption{Efficient sampling in the Morris method ($n=2,~ p=4,~ r=3,~ s=1$): random sampling results in 12 model evaluations (left); this number can be reduced to 9 by forming elementary paths (right).}%
\label{fig:eff_vs_paths}%
\end{figure}%

The analysis described above implies performing a total of $2n \cdot r$ model evaluations. \cite{morris} proposed an efficient sampling scheme. It relies on elementary effects that share endpoints on the latin hypercube grid (Fig.~\ref{fig:elem_paths}), effectively leading to $(n+1) \cdot r$ \emph{elementary paths}. Such a path starts at a random position on the grid and sequentially travels one step of length $\delta$ over each dimension. This effectively reduces the number of required model evaluations by a factor of $2$.

The choices for $p$, $r$ and $s$ have a significant impact on the outcome of the sensitivity analysis. If a high value of $p$ is considered, which means that a high number of levels will be partitioned, one may think that the accuracy of the sampling has been increased. However, if this is not related to a high value of $r$, many of the levels will remain unexplored. Also, the value of $s$ depends on the choice of $p$. According to \cite{morris}, a convenient choice is $s = \frac{p}{2}$ (assuming $p$ is even), while previous studies \citep{campolongo2007} have demonstrated that  $p=4$ and $r=10$ produce valuable results in many cases.

\change{A short introduction to copulas is presented in the next section.}

\begin{figure}%
\centering%
\includegraphics[width=\columnwidth]{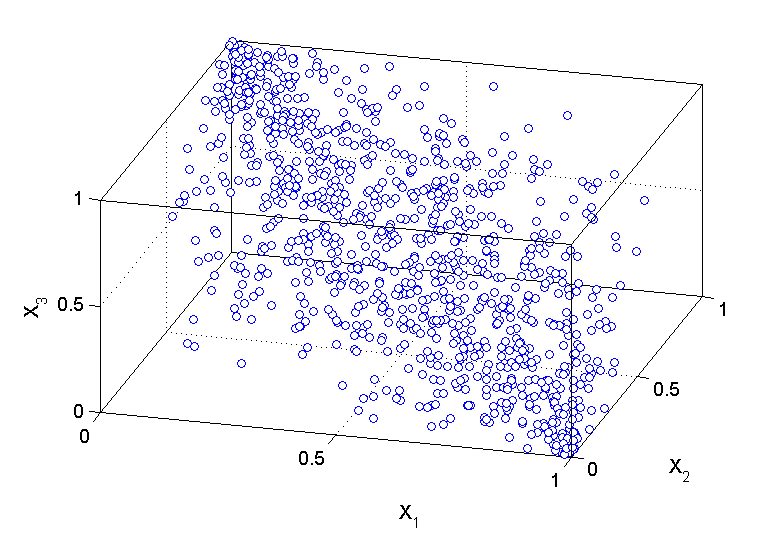}%
\caption{\change{Scatter plot of 500 samples from a Gaussian copula with correlations $\rho(x_1,x_2)=\rho(x_1,x_3)=-0.7$ and $\rho(x_2,x_3)=0.7$.}}%
\label{fig:scatter}%
\end{figure}

\change{\section{Copulas}
\label{sec:copula}

A copula is a joint distribution, defined on an $n$-dimensional unit hypercube with uniform marginal distributions \citep{nelsen2007}. It is a very popular way of representing the joint distribution, since it separates the influence of marginal distributions from the influence of parameter dependencies.

The joint cumulative distribution function $F(x_1,...,x_n)$ of random variables $X=(x_1,\dots,x_n)$ with
marginal distributions denoted as $F_i(x_i), i=1,...,n$ can be represented with copula $C$ as follows,
\begin{equation}
F(x_1,...,x_n) = C(F_1(x_1),...,F_n(x_n)),
\end{equation}
which is unique if $(x_1,\dots,x_n)$ are continuous \citep{nelsen2007}. 

The most popular copulas used in practice are Gaussian, Student-t and copulas from the Archimedean family. In Figure~\ref{fig:scatter}, a scatter plot of samples from a three dimensional Gaussian copula with correlations $\rho(x_1,x_2)=\rho(x_1,x_3)=-0.7$ and $\rho(x_2,x_3)=0.7$ is presented. The larger concentration of points close to the (1,0,0) and (0,1,1) is due to to the negative correlation between the first parameter and the remaining two.
}


\section{Copula-based Morris method}
\label{sec:extended_morris}

The Morris method \citep{morris} was conceptually designed for models with independent parameters. However, most often,  model parameters are related to each other; disregarding this association results in an invalid description of the physical system. Sensitivity analysis based on independent random sampling, as is the one performed by the classic Morris method, is not applicable in these cases, since it breaks the underlying model assumptions, possibly leading to unrealistic behavior. This has motivated the need to develop a general method for sensitivity analysis. For this reason, this section introduces a novel copula-based approach, able to account for a wide range of dependencies between the model parameters.

\begin{figure}%
\centering%
\subfloat[the whole grid]{\includegraphics[width=\columnwidth]{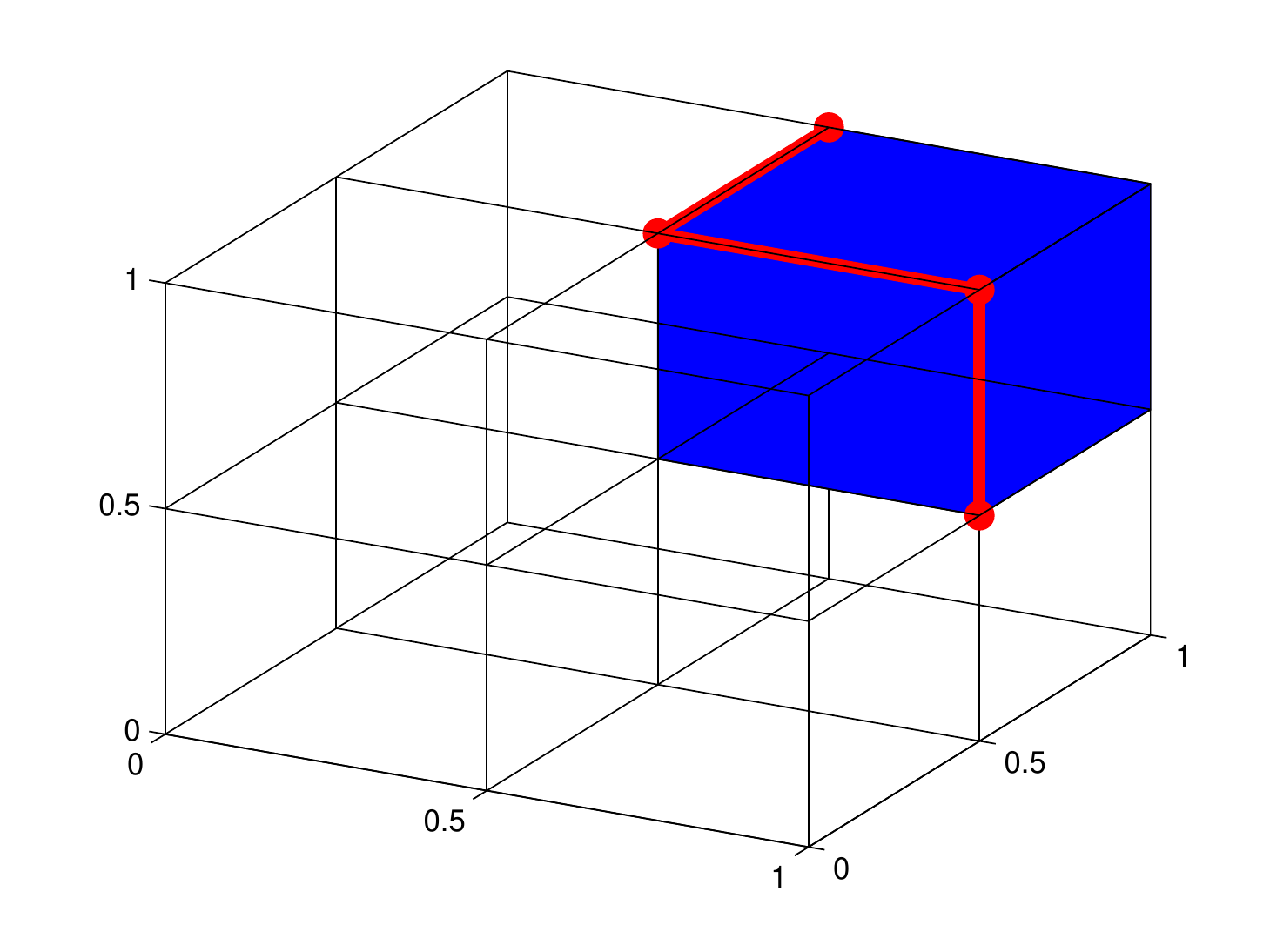}\label{fig:path_cell1}}\\%
\subfloat[zoom on the cell]{\includegraphics[width=\columnwidth]{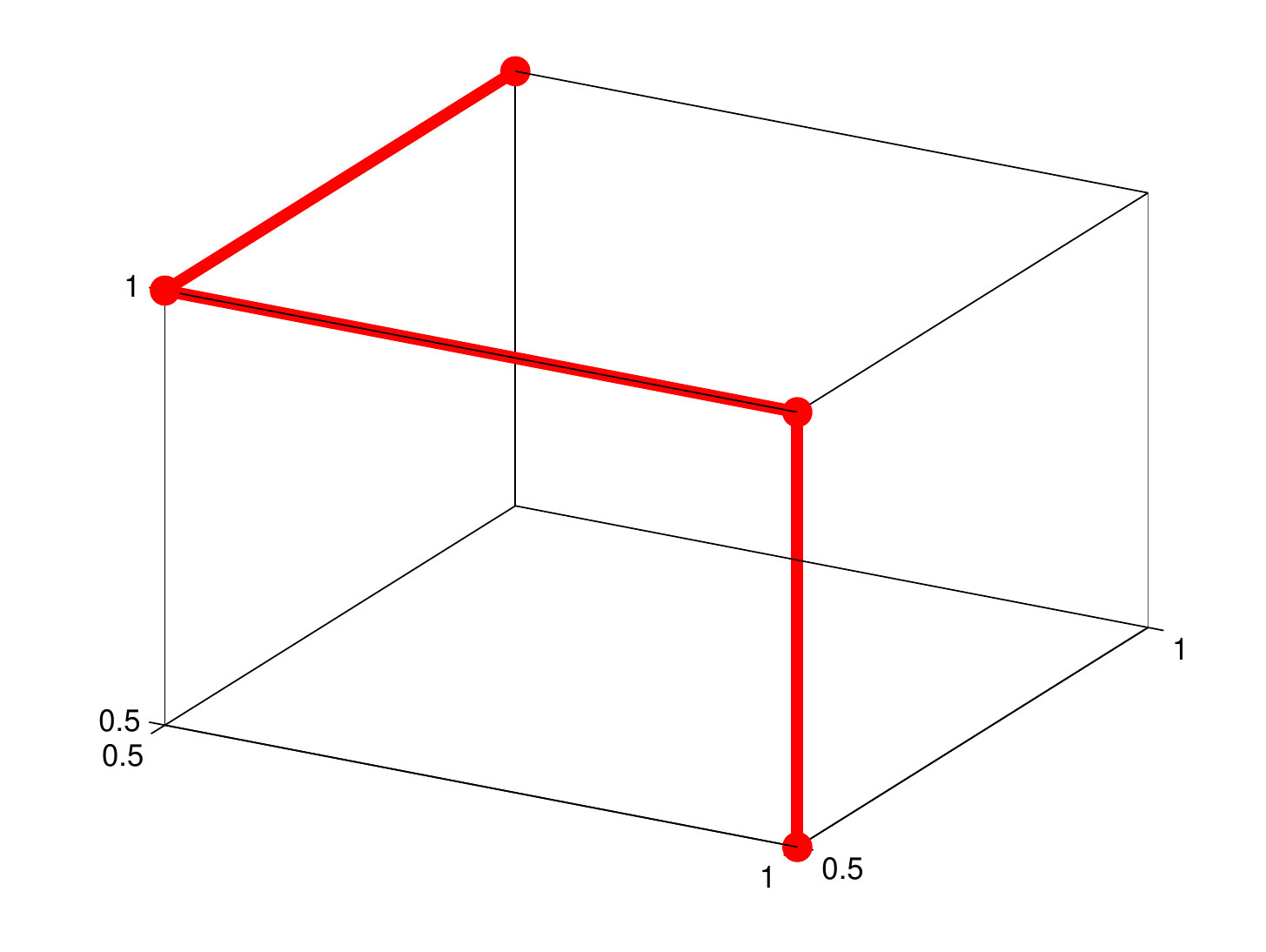}\label{fig:path_cell2}}%
\caption{Geometric reinterpretation of an elementary path ($n=3,~ p=3,~ s=1$).}%
\label{fig:path_cell}%
\end{figure}

As discussed before, the elementary paths are the building blocks of the Morris method. Without loss of generality, consider the case when the Morris step is equal to one cell, i.e. $\delta = \frac{1}{p-1}$. Then, as illustrated in Fig.~\ref{fig:path_cell}, each path runs on the contour of a grid cell, starting in one of its corners and ending in the opposite one (since all coordinates are successively altered with $\pm \delta$).

The copula-based method relies on the key observation that the sampling of a path can be done, equivalently, in the following three steps:
\begin{enumerate}%
\item \textit{Choosing the target grid block}
\item \textit{Choosing the starting point} as one of the corners of the grid block
\item \textit{Choosing the traversal order} of the contour segments, in order to reach the opposite corner
\end{enumerate}

For example, the path in Figure~\ref{fig:path_cell} was obtained by first choosing the blue-shaded grid cell, then its lower-right corner as the starting point, A. In order to calculate the elementary effects, a path must be chosen such that all the parameters, three in this case, are varied, one at a time, with $\delta$. Note that there are $3! = 6$ different ways of traversing this grid cell from A to B. In this case, parameter $x_3$ is changed first, followed by $x_1$ and finally, $x_2$. Thus, determining an order of traversal is equivalent to choosing a permutation of the set $\{1,2,3\}$.

Note that traversing a path in reverse (from B to A) does not produce new results, since it decomposes into the same elementary effects. Therefore, there are two different ways to sample the same path: choosing its start corner and corresponding permutation $\mathbf{\pi}$, or choosing its end corner and the reverse of permutation $\mathbf{\pi}$. Since this is true for all elementary paths, their probability of being selected remains uniformly distributed (in accordance to the classic formulation in \cite{morris}), without any alteration of the sampling strategy.

\begin{figure}%
\centering%
\subfloat[the whole grid]{\includegraphics[width=\columnwidth]{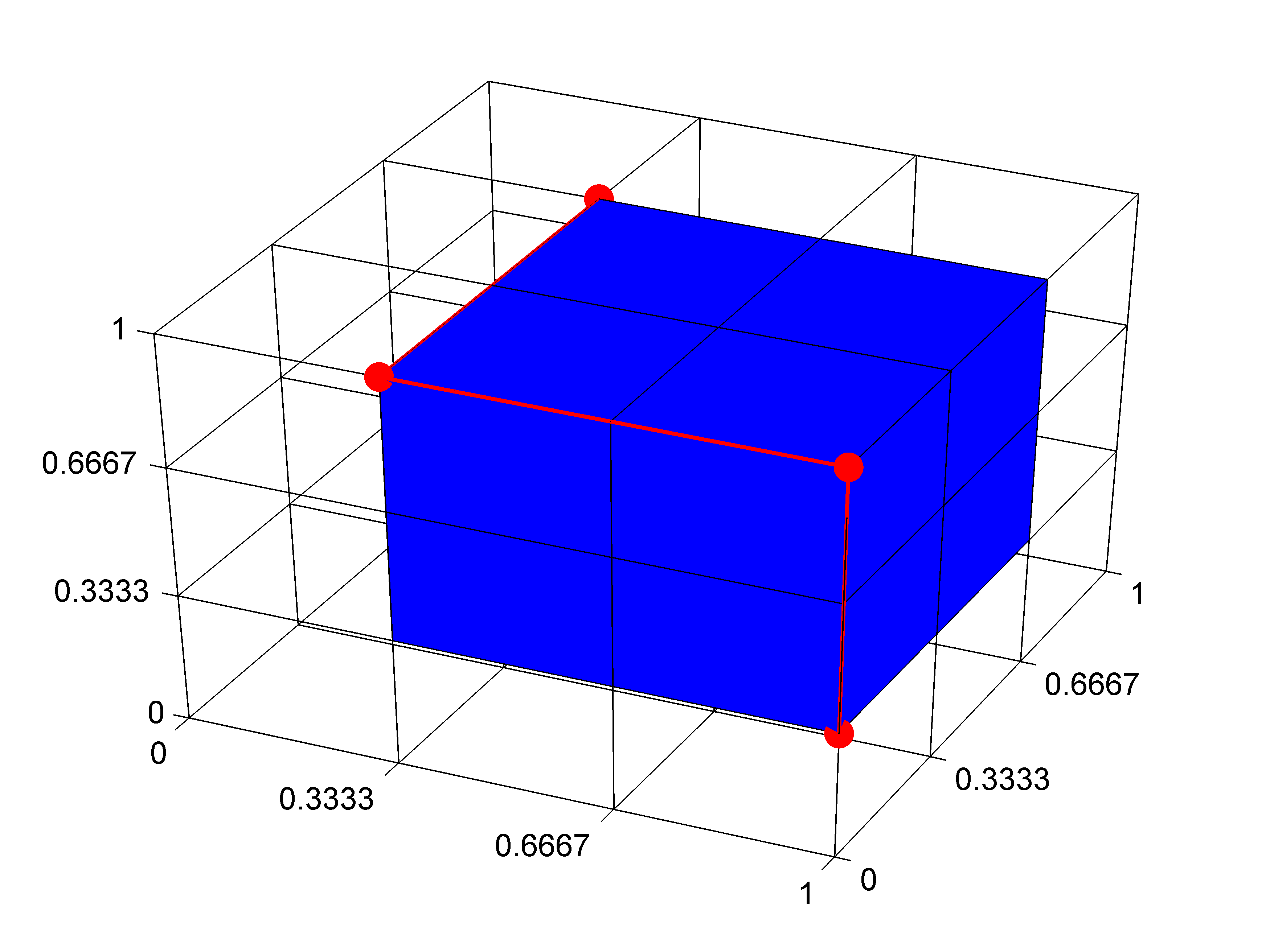}\label{fig:path_block1l}}\\%
\subfloat[zoom on the block]{\includegraphics[width=\columnwidth]{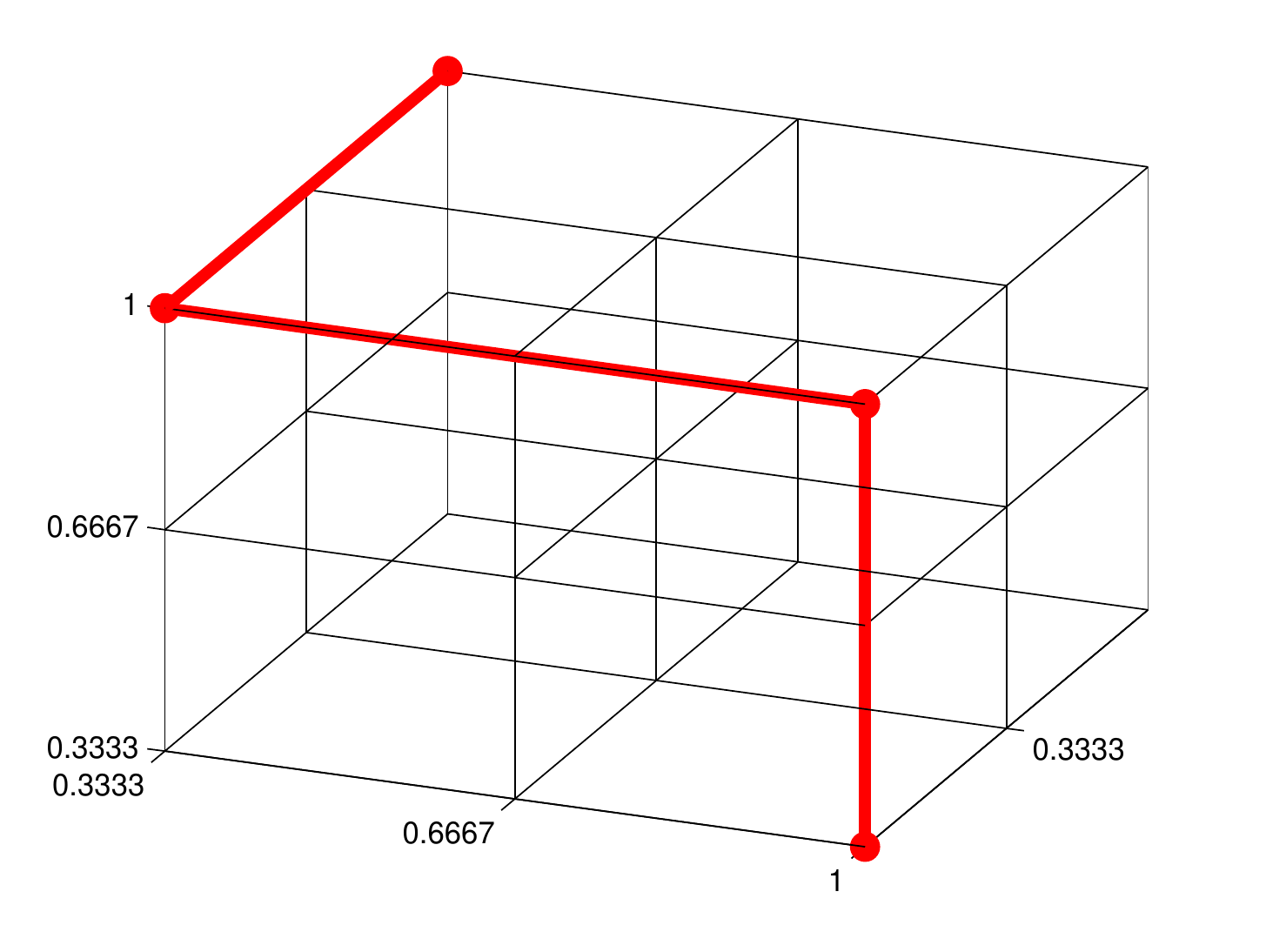}\label{fig:path_block2}}%
\caption{Elementary path for $n=3,~ p=4,~ s=2$.}%
\label{fig:path_block}%
\end{figure}

If the Morris step is higher than one grid cell \eqref{eq:step}, the only difference is that the path is drawn on the contour of a $s \times s$ grid block (Figure~\ref{fig:path_block}). Note that, even though neighboring blocks intersect each other, they spawn different elementary paths and, hence, are \emph{conceptually} disjunct.

This geometric interpretation allows us to compute the total number of possible paths on the unit hypercube as:
\begin{equation}
\begin{split}
& N_{cells} = (p-s)^n, \hspace{0.75cm} N_{corners} = 2^n, \hspace{0.75cm} N_{orders} = n!,  \\
& N_{paths} = N_{cells} \cdot N_{corners} \cdot N_{orders}~ /~ 2
\end{split}
\end{equation}
where $n$ is the number of parameters, $p$ is the number of discretization levels and $s$ is the Morris step size. More importantly, sampling dependence constraints can now be introduced into each of the three steps enumerated above, \change{by appropriately altering the sampling probabilities of the elementary effects}.

\subsection{Choosing the target grid block}
\label{sec:step1}
The position of the grid block containing an elementary path gives the range of values within which the parameters are varied sequentially to compute elementary effects. Previous studies state that having the paths sufficiently spread within the unit hypercube is vital for the results of the analysis. For this purpose, \cite{campolongo2007} introduce a penalty term based on Euclidean distances, while \cite{vanGriensven2006} use Latin Hypercube Sampling \citep[LHS, see][]{mcKay}, instead of Monte-Carlo.

\begin{figure}%
\centering%
\subfloat[Copula samples]{\includegraphics[width=\columnwidth]{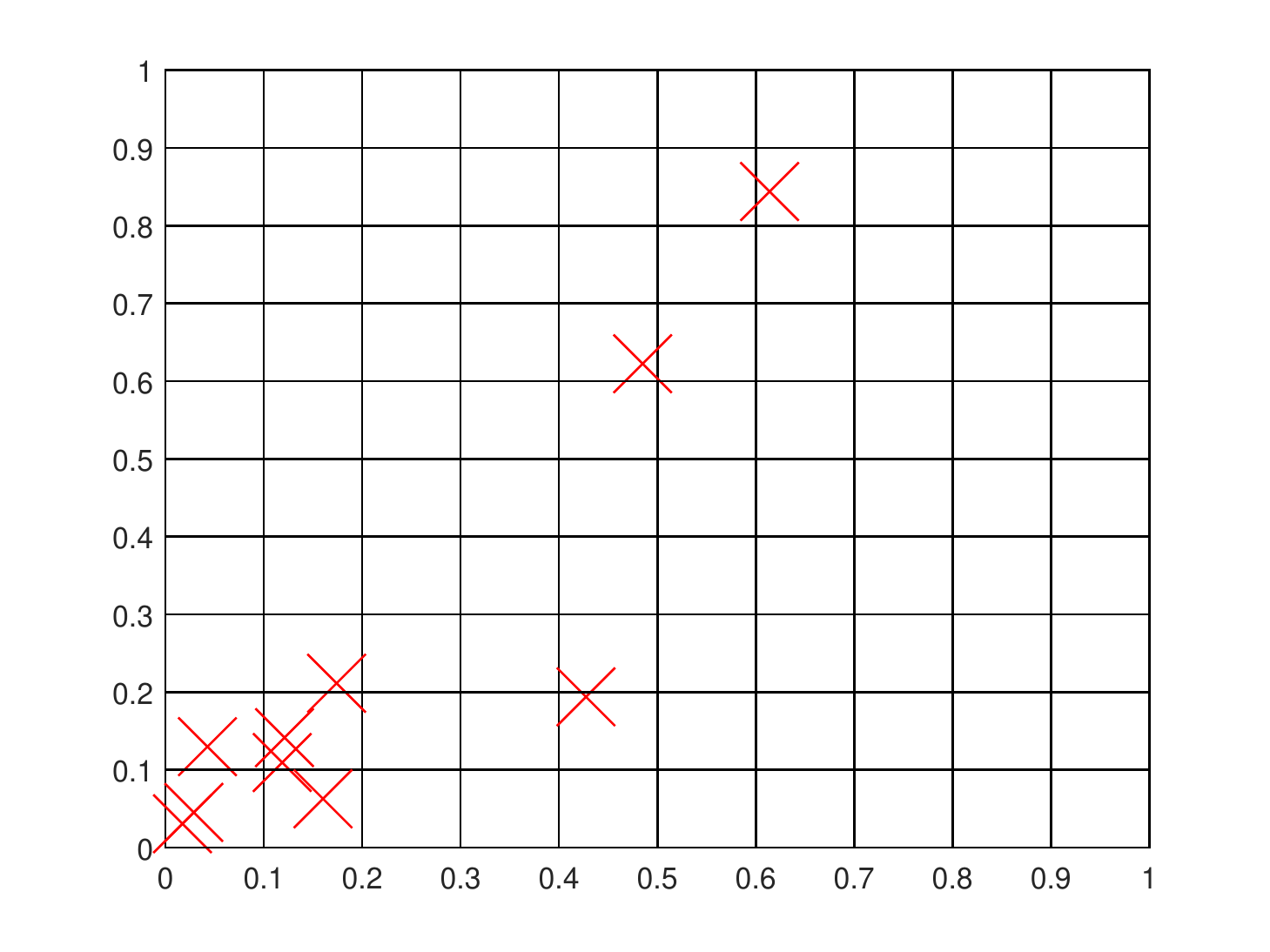}\label{fig:lhs}}\\%
\subfloat[Latin hypercube reordered samples]{\includegraphics[width=\columnwidth]{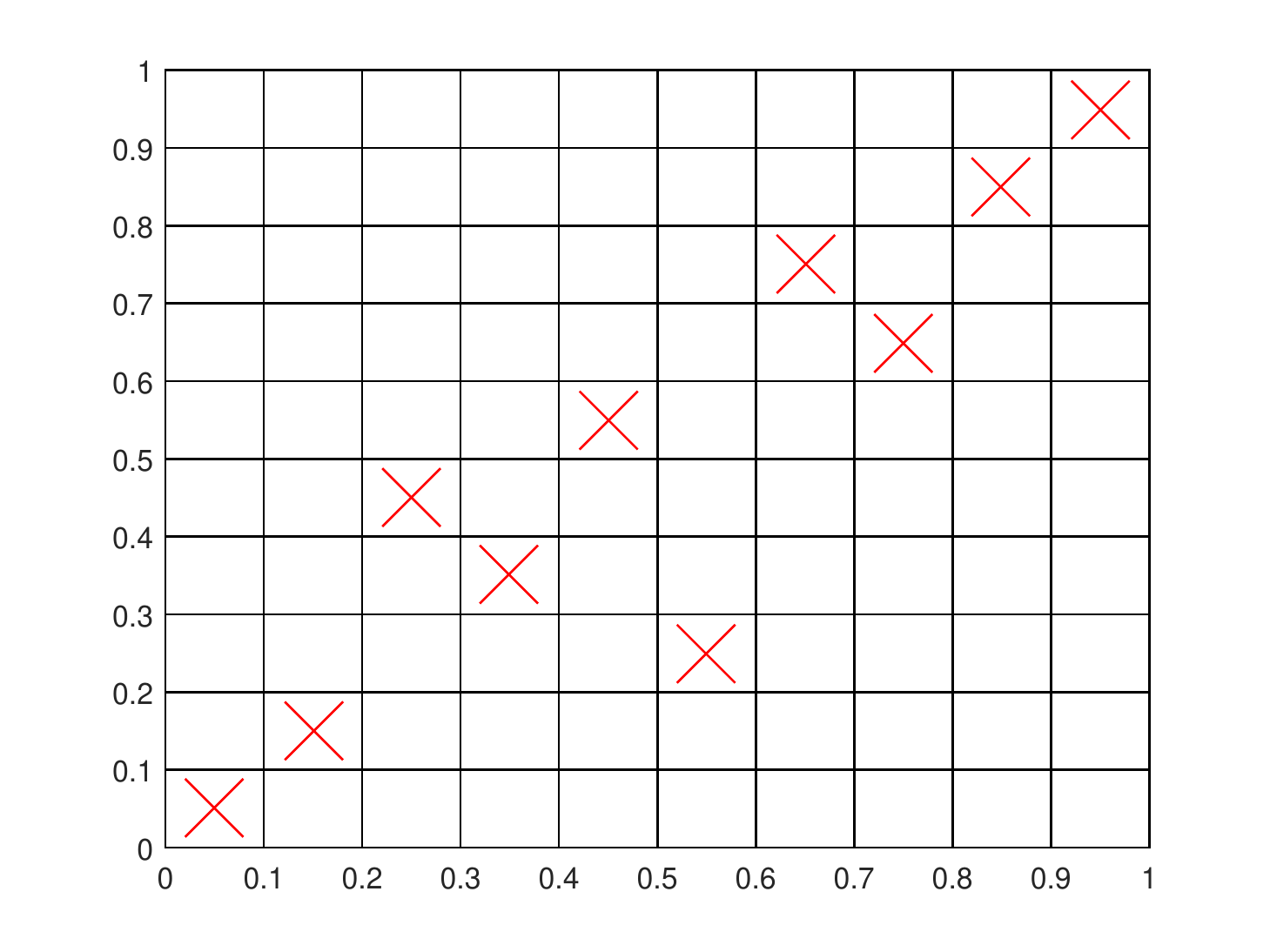}\label{fig:lhsd}}%
\caption{Using LHSD to ensure an even spread of the copula samples within the parameter space ($n=2,~ p=11,~ s=1$).}%
\label{fig:lhs_vs_lhsd}%
\end{figure}%

\change{The goal of the new method is to constrain the sampling of the blocks in accordance to the available information about parameter dependencies. To this end, the first step is to specify a copula \citep{nelsen2007} which captures these dependencies. As presented in Section \ref{sec:copula} one can extract a copula from the joint distribution between model parameters by transforming the margins to be uniform on (0,1). This can be done simply by linear scaling if the parameters are uniformly distributed over their original ranges, otherwise marginal distributions need to be applied. Latin hypercube sampling is then performed on the copula,thus ensuring a good coverage of the parameter space.
\footnote{\change{This part of the algorithm could be done differently if only information about correlations between parameters was available. One could first obtain $l$ Latin hypercube samples as specified in \citep{mcKay} and impose correlation constraints with the  \cite{iman} method. However, since that method uses van der Warden scores (based on the normal distribution) that are linearly transformed with a lower triangular matrix obtained from the desired correlation matrix, it is approximately equivalent with the method presented in this paper, while employing a normal copula.}}

For example, in $n=2$ dimensions, there will be exactly one sample in each row and each column (compare Figure~\ref{fig:lhs} with Figure~\ref{fig:lhsd}) as in the original latin hypercube sampling method \citep{mcKay}, with the added effect of preserving dependencies between parameters, due to the copula.}  The algorithm used to achieve this is Latin Hypercube Sampling with Dependence (LHSD) and has been recently proposed by \cite{packham}. Formally, considering a hypercube of $l^n$ grid cells, LHSD operates by taking $l$ samples from the copula, $u^{(1)}, \dots, u^{(l)} \in \mathbb{R}^n$ (as in Figure~\ref{fig:lhs}) and arranges them to get one sample in each row and column, while preserving their ranking. \change{More precisely,the rank statistics of the $i$-th sample of parameter $j$ are computed as
\begin{equation}%
R_j[i] = \sum_{k=1}^l \mathbb{1}_{\{u^{(k)}[j] \leq u^{(i)}[j]\}} \hspace{0.6cm} i=1,\dots,l \hspace{0.5cm} j=1,\dots,n%
\label{eq:rank_stat}%
\end{equation}%
where $\mathbb{1}_{S}$ denotes the indicator function of set $S$.} $R_j[i]$ effectively represents the order of the sample in $(u^{(1)}[j],...,u^{(l)}[j])$.

Finally, the vector containing the coordinates of the \change{origin of the target cell (i.e. its lower-left corner)} is determined as:
\begin{equation}%
\mathbf{cell}^{(i)}[j] = \frac{R_j[i] - 1}{l} \hspace{1.25cm} i=1,\dots,l \hspace{0.5cm} j=1,\dots,n%
\label{eq:lhsd_sampling}%
\end{equation}%

Note that, by the nature of LHSD, the number of samples needs to be a multiple the size of the hypercube. However, sensitivity studies may require an arbitrary number of samples. To maintain the flexibility of the Morris method, the LHSD algorithm can be repeated several times, until there is a sufficient number of samples (the excess can be discarded).


\subsection{Choosing the starting point}
\label{sec:step2}

\begin{figure}%
\centering%
\includegraphics[width=\columnwidth]{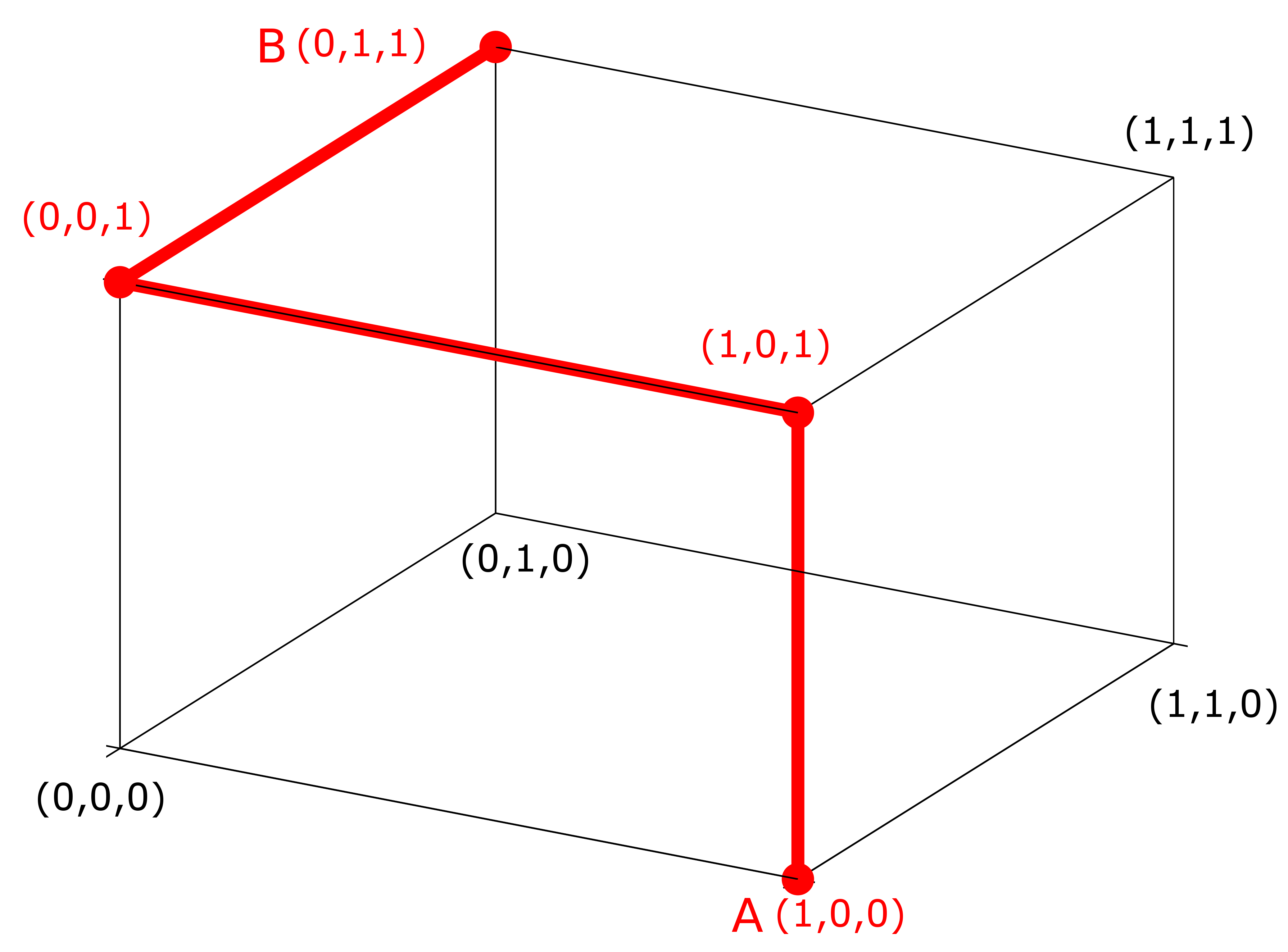}%
\caption{Binary representation of the nodes on the path, starting in corner A (i.e. $\mathbf{b}_s = (1,0,0)$) and ending in B (i.e. $\mathbf{b}_e = (0,1,1)$).}%
\label{fig:binary_cube}%
\end{figure}

\change{For each sampled grid block, $\mathbf{cell}^{(i)}$, the starting corner of the path is randomly sampled such that the dependence information between parameters is preserved. The idea is simple -- each corner in the grid is treated as a realization of an $n$-dimensional discrete distribution with $p$ possible values, namely $\{0,1,2,...,p-1\}$, for each factor.

The advantage of using a copula is that the marginal distributions of the factors are removed through marginal transformation or linear scaling. The marginal probability of each factor $x_i$ taking value $j\in \{0,1,2,...,p-1\}$ is $\text{Prob}(x_i=j)=\frac{1}{p}$. As such, the finite difference formula presented in \cite{nelsen2007} can be used to compute the probability of each point specified on the grid,
\begin{equation}%
\text{Prob}(X_1=j_1,\dots,X_n=j_n)=\Delta_{a_1}^{b_1}\Delta_{a_2}^{b_2} \dots \Delta_{a_n}^{b_n}C,%
\end{equation}%
where $a_i=\frac{j_i}{p}$, $b_i=\frac{j_i+1}{p}$ and
\begin{equation}%
\begin{aligned}%
\Delta_{a_k}^{b_k}C~ =~ & C(u_1,\dots,u_{k-1}, b_{k}, u_{k+1},\dots,u_n)~  - \\
&  C(u_1,\dots,u_{k-1}, a_{k}, u_{k+1},\dots,u_n)%
\end{aligned}%
\end{equation}%
Hence when $n=3$
\begin{equation}%
\begin{aligned}%
&\text{Prob}(X_1=x_1,X_2=x_2,X_3=x_3)= \\%
&C(b_1,b_2,b_3)-C(a_1,b_2,b_3)-C(b_1,a_2,b_3)+C(a_1,a_2,b_3) \\%
-&C(b_1,b_2,a_3)+C(a_1,b_2,a_3)+C(b_1,a_2,a_3)-C(a_1,a_2,a_3).%
\end{aligned}%
\label{eq:prob}%
\end{equation}%

Consider the formula~\ref{eq:prob} when $p=2$. In this case the hypercube is composed of only one cell with eight corners. Each factor can take only two possible values $x_i=\{0,1\}$. $p_{0,0,0}=\text{Prob}(X_1=0,X_2=0,X_3=0)=C(\frac{1}{2},\frac{1}{2},\frac{1}{2})$, with $a_i=0, b_i=\frac{1}{2}, i=1,2,3$, since any copula evaluated at point zero is 0. Using a normal copula with the correlation matrix presented in Section \ref{sec:copula}, $p_{0,0,0}=0.0633$. Similarly, $p_{0,1,0}=p_{0,0,1}$$=p_{1,1,0}=p_{1,0,1}=p_{1,1,1}=0.0633$. However, the probability of the point (0,0,1) (as well as (1,1,0)) is much higher (see Figure~\ref{fig:scatter}) and it can be calculated  as $p_{1,0,0}=C(1,\frac{1}{2},\frac{1}{2})-C(\frac{1}{2},\frac{1}{2},\frac{1}{2})=0.3101$ with $a_i=0, b_i=\frac{1}{2}, i=2,3$ and $a_1=\frac{1}{2}, b_1=1$. The starting point is sampled according to the calculated distribution, hence there is a much larger chance to choose point (1,0,0) or (0,1,1) over the other options.

This procedure can be used to compute distributions of corner points for each cell on the grid. However, for large $p$ it can prove to be a computationally demanding task. It is only of interest to do this for the grid cells resulting from LHSD, however If this is still a large number, one can compute the distribution only once, as illustrated in the example above for $p=2$, and assume that it applies to all grid cells. In the case of the Gaussian copula, this simplified procedure would be sufficient. For more complicated copulas, i.e. with asymmetries and tail dependencies, however, the assumption would not hold.

Finally, after the starting corner is determined, the corresponding elementary path will end in the opposite corner, since it is composed of one elementary effect for each parameter. For convenience, each of the grid cell's corners are assigned a binary representation, starting with (0,0,0) in the origin (see Fig.~\ref{fig:binary_cube}). Then, the end point is determined by negating the representation of the sampled starting point. Note that there are $2^n$  possible paths between these points -- the choice between them is explained in the next paragraph.}

\subsection{Choosing the traversal order}
\label{sec:step3}

The order of traversal is given by a randomly sampled permutation $\mathbf{\pi}^{(i)}$ which describes the way to get from the starting point to the opposite corner by changing one factor at the time. The path's vertices are determined by sequentially negating the components in the starting point's binary representation \change{(similar to a Gray code sequence, obtained by swapping one bit at a time)}. For example, the path in Figure~\ref{fig:binary_cube} was obtained using the permutation $\{3,1,2\}$, corresponding to the change along the $\{z,x,y\}$ axes:

\begin{center}
\mbox{\bordermatrix{~ & x & y & z \cr%
\textbf{b}_s \rightarrow & 1 & 0 & \textbf{0} \cr%
& \textbf{1} & 0 & \textbf{1} \cr%
& \textbf{0} & \textbf{0} & 1 \cr%
\textbf{b}_e \rightarrow  & 0 & \textbf{1} & 1}}%
\end{center}


\subsection{Method summary}
To conclude, the copula-based Morris method follows the following outline:\\~\\
\textit{Prerequisites}
\begin{itemize}
\item A model that takes $n$ parameters, $M(x_1,\dots,x_n)$, with their corresponding ranges.
\item A copula $C$ that best describes the dependence between the parameters. In the absence of any prior information the independence copula can be assumed, whereas, if there are known correlations between the parameters, $\rho_{i,j}$, then a Gaussian copula is appropriate. For more complex dependency structures (e.g. tail dependence), one is free to use a copula from the Archimedean family (Clayton, Gumbell, etc.) or infer an empirical copula from a pre-existing set of model runs.
\item The number of levels, $p$, and step size, $s$, for the Morris method.
\item The number of desired paths, $r$, by taking into account that $(n+1) \times r$ model runs are necessary.
\end{itemize}
\textit{Algorithm}
\begin{enumerate}
\item Define the grid as a $p$-level $n$-dimensional unit hypercube.
\item Sample $r$ vectors, $u_i$, from the copula $C$.
\item Compute the rank statistics \eqref{eq:rank_stat}.
\item Compute the LHSD samples \eqref{eq:lhsd_sampling}, which represent the grid blocks.
\item For each grid block, determine the start and corresponding end point, as explained in paragraph \ref{sec:step2}.
\item Determine the order of traversal of the path's segments by sampling a permutation, $\mathbf{\pi}^{(i)}$, and determine the path, as explained in paragraph \ref{sec:step3}.
\item Evaluate the model at each point along the paths and compute the elementary effects \eqref{eq:elem_effect}.
\item Compute and interpret the sensitivity measures $\mu_i$, ${\mu_i}^*$ and $\sigma_i$ using \eqref{eq:measure1} and \eqref{eq:measure2}.
\end{enumerate}


\section{Sensitivity analysis results}

\begin{figure*}[htb!]%
\centering%
\includegraphics[width=\textwidth]{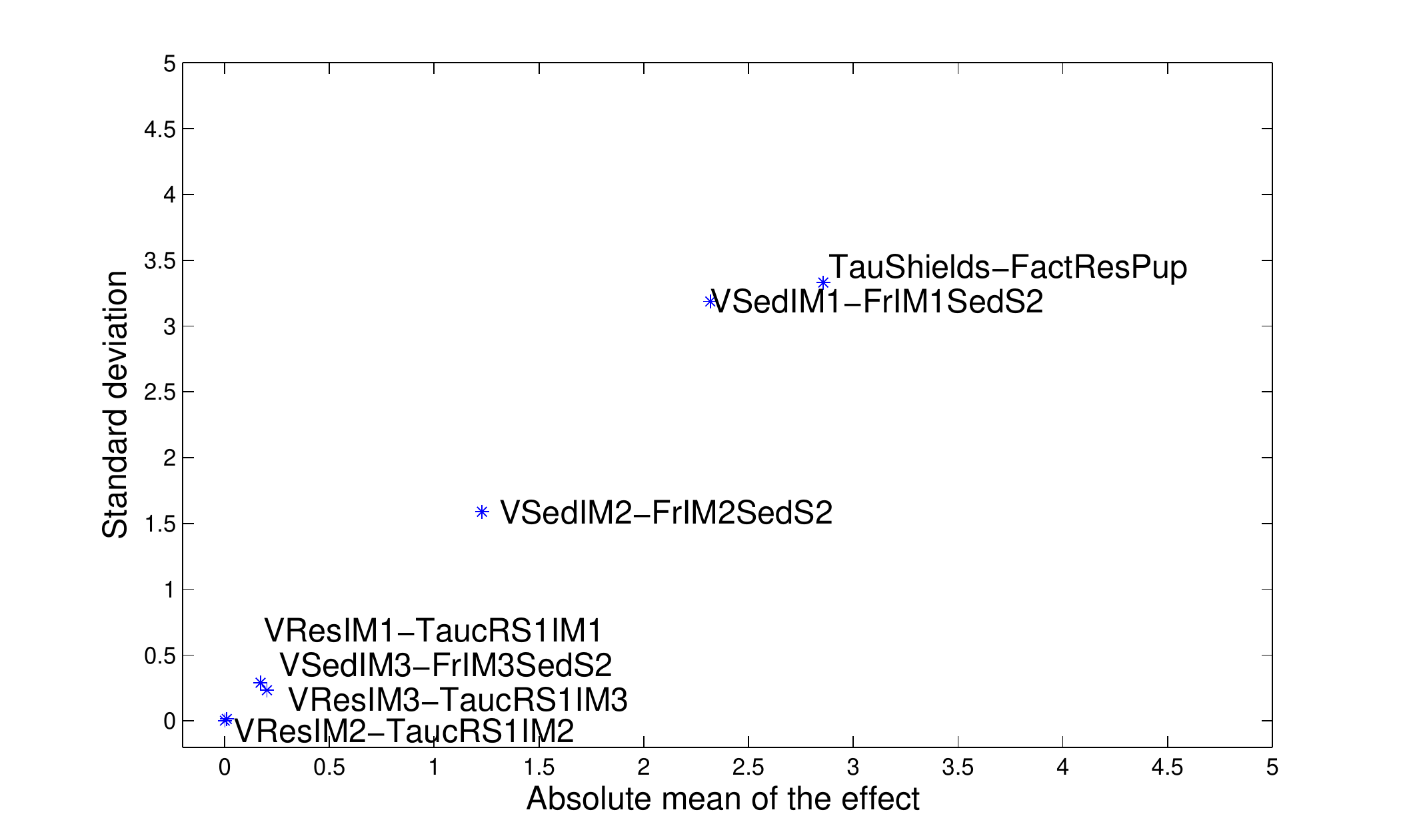}%
\caption{Copula-based Morris method sensitivity results.}%
\label{fig:resultsCop}%
\includegraphics[width=\textwidth]{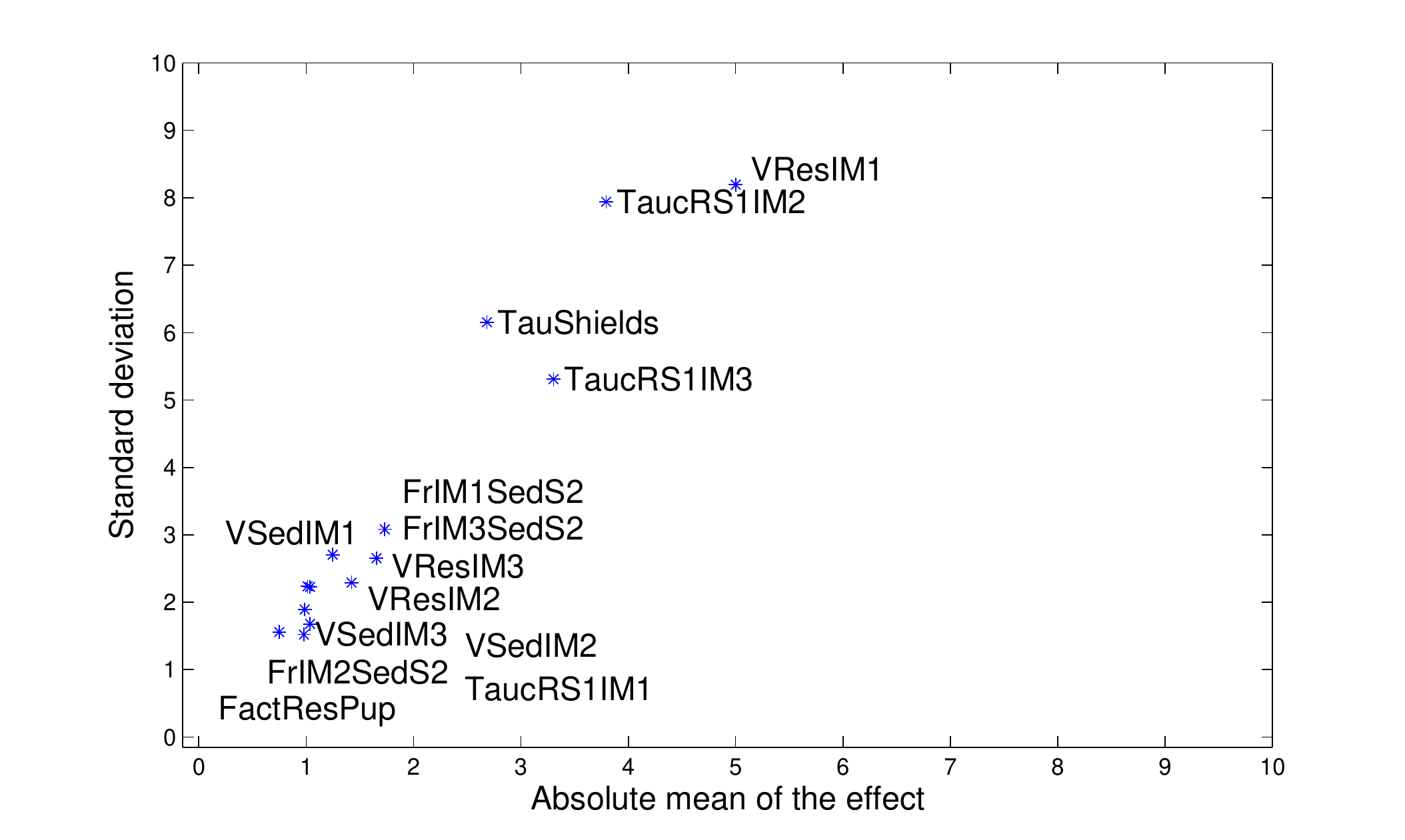}%
\caption{Classic Morris method sensitivity results.}%
\label{fig:resultsIndep}%
\end{figure*}%

The methodology proposed in section \ref{sec:extended_morris} has been applied to the Delft3D-WAQ model (described in Section~\ref{sec:delft3d}). Recall that, in order to respect the interactions between the parameters, we separate them into 7 perfectly correlated pairs. This leads to the construction of a Gaussian copula with the rank-correlations given in Table~\ref{tab:paramsCor}, which enables the computation of cumulative elementary effects of each pair, rather than that of each individual parameter. \change{Due to this specific choice of correlation structure, the sampling will favor grid cells lying on the hypercube's diagonal (anti-diagonal) for factors completely positively (negatively) correlated. Subsequently, the same dependencies are used to constrain the choice of starting points for the elementary paths, while the order of traversal is randomly sampled.}

Since the Delft3D-WAQ model is computationally expensive (3 hours run time on a coarse grid and 11 hours for a fine grid),  the number of simulations that can be performed for sensitivity analysis is limited. Therefore, the parameter space (unit hypercube) was divided into $p=4$ equidistant levels, on which $r=10$ elementary paths were sampled with a Morris step of $s=2$ cells. Therefore, a total number of $80$ simulations were performed for the sensitivity study.

For the sake of comparison, a separate set of simulations was performed, where the parameters were sampled using Morris' classic algorithm (thus, assuming complete independence). The comparative results are depicted in Fig. ~\ref{fig:resultsCop} and in Fig.~\ref{fig:resultsIndep} and detailed in Tables~\ref{tab:measuresCop} and \ref{tab:measuresIndep}.

\begin{table}[htb!]%
\centering%
\change{\caption{Sensitivity measures for the correlated pairs, using the copula-based approach. The ranking is done ordered in decreasing order of $\mu^*$.\label{tab:measuresCop}}}%
\begin{tabular}[t]{ C{4cm} c c c } \midrule%
\bf{Pair}		& $\mathbf{\mu}$		& $\mathbf{\mu^{*}}$ & $\mathbf{\sigma}$ \\ \midrule%
\change{$\tau_\text{Shields}$ -- $Fact_\text{res Pup}$}					& 0.023		& 2.857	& 3.331 \\%
\change{$V_{\text{sed IM}_1}$ -- $Fr_{\text{IM}_1\text{ sed S}_2}$}		& -0.077		& 2.317	& 3.187 \\%
\change{$V_{\text{sed IM}_2}$ -- $Fr_{\text{IM}_2\text{ sed S}_2}$}		&  0.496   	& 1.228	& 1.589 \\%
\change{$V_{\text{sed IM}_3}$ -- $Fr_{\text{IM}_3\text{ sed S}_2}$}		&  0.063		& 0.202	& 0.233 \\%
\change{$V_{\text{res IM}_1}$ -- $\tau_{\text{cr S}_1\text{ IM}_1}$}		&  0.019		& 0.171	& 0.290 \\%
\change{$V_{\text{res IM}_2}$ -- $\tau_{\text{cr S}_1\text{ IM}_2}$}		&  0.003		& 0.011	& 0.015 \\%
\change{$V_{\text{res IM}_3}$ -- $\tau_{\text{cr S}_1\text{ IM}_3}$}		&  0.000		& 0.001	& 0.002 \\ \midrule%
\end{tabular}%
\end{table}%

\begin{table}[htb!]%
\centering%
\change{\caption{Sensitivity measures for each parameter individually, using the classic Morris method, which does not support dependencies. The ranking is done ordered in decreasing order of $\mu^*$.\label{tab:measuresIndep}}}%
\begin{tabular}[t]{ C{4cm} c c c } \midrule%
\bf{Parameter}		& $\mathbf{\mu}$		& $\mathbf{\mu^{*}}$		& $\mathbf{\sigma}$ \\ \midrule%
\change{$V_{\text{res IM}_1}$}    				& -0.160  & 5.002     & 8.193 \\%
\change{$\tau_{\text{cr S}_1\text{ IM}_2}$}	& -1.666  & 3.794     & 7.939 \\%
\change{$\tau_{\text{cr S}_1\text{ IM}_3}$}	&  2.815  & 3.304     & 5.311 \\%
\change{$\tau_\text{Shields}$}				& -2.037  & 2.684     & 6.153 \\%
\change{$Fr_{\text{IM}_1\text{ sed S}_2}$}	& -0.546  & 1.731     & 3.083 \\%
\change{$V_{\text{res IM}_3}$}				& -1.221  & 1.655     & 2.652 \\%
\change{$V_{\text{res IM}_2}$}				& -0.412  & 1.423     & 2.291 \\%
\change{$Fr_{\text{IM}_3\text{ sed S}_2}$}	& -0.450  & 1.247     & 2.706 \\%
\change{$V_{\text{sed IM}_1}$}				& -0.732  & 1.037     & 2.228 \\%
\change{$\tau_{\text{cr S}_1\text{ IM}_1}$}	&  0.660  & 1.037     & 1.678 \\%
\change{$V_{\text{sed IM}_2}$}				&  0.918  & 1.013     & 2.235 \\%
\change{$Fr_{\text{IM}_2\text{ sed S}_2}$}	&  0.483  & 0.986     & 1.894 \\%
\change{$V_{\text{sed IM}_3}$}				&  0.642  & 0.979     & 1.523 \\%
\change{$Fact_\text{res Pup}$}				& -0.251 & 0.749     & 1.559 \\ \midrule%
\end{tabular}%
\end{table}%

The results of the copula-based Morris method match the expectations induced by the physics of the system and defined during the expert judgment exercise. The \change{parameters to which the model is most sensitive to (presented in pairs)} are, in this order:
\begin{enumerate}%
\item \change{$\tau_\text{Shields}$ -- $Fact_\text{res Pup}$}%
\item \change{$V_{\text{sed IM}_1}$ -- $Fr_{\text{IM}_1\text{ sed S}_2}$}%
\item \change{$V_{\text{sed IM}_2}$ -- $Fr_{\text{IM}_2\text{ sed S}_2}$}%
\end{enumerate}%

As seen in Table \ref{tab:measuresCop}, the values of $\mu$ and $\mu^{*}$ for these parameter pairs differ significantly, which suggests a high interaction with the other pairs. The pair \change{$\tau_\text{Shields}$ -- $Fact_\text{res Pup}$} is mainly responsible for the sand resuspension processes from the second bed layer releasing silt during high stress events (e.g. high waves, spring tides) while the pairs \change{$V_{\text{sed IM}_i}$ -- $Fr_{\text{IM}_i\text{ sed S}_2}$}, $i \in \{1,2\}$, are involved in the deposition processes of the medium and coarse particles from the water column into the two bed layers. Note that the pairs \change{$V_{\text{res IM}_i}$ -- $\tau_{\text{cr S}_1\text{ IM}_i}$}, $i \in \{1,2,3\}$, which are involved in the resuspension process from the fluffy bed layer by weaker stress conditions (e.g. semi-diurnal tidal fluctuations), are of less impact on the model output variability. From this set, the resuspension for the medium size particles ($\text{IM}_1$) has the highest impact.

On the other hand, the results of the classic Morris method rank the first-order resuspension rate for medium particles \change{$V_{\text{res IM}_1}$}, the critical resuspension stress from the layer $S_1$ for the coarse and fine particles \change{$\tau_{\text{cr S}_1\text{ IM}_i}$}, $i \in \{2,3\}$, and the critical shear stress \change{$\tau_\text{Shields}$} as the top four most influential parameters. \change{$\tau_\text{Shields}$} appears in both rankings as a \change{sensitive} parameter.

The comparison shows that, under the assumption of independence, the dominant process is the resuspension from layer $S_1$ followed by the resuspension from the layer $S_2$, while under the copula-based approach, the dominant process is the resuspension from the second layer succeeded by deposition. In the model setup, $S_1$ represents a thin fluff\change{y} layer consisting of rapidly eroding mud, while most sediment is stored in the sandy layer $S_2$. When the bed shear stress \change{$\tau_\text{Shields}$} exceeds a critical value (energetic conditions such as spring tides or storms) the sandy layer becomes mobile and the sediment is released in the water column. It is therefore expected that the total SPM concentration in the water column increases significantly. On the other side, during calm conditions, the presence of sediment in the water column is influenced by the deposition rates. As such, the results of the copula-based sensitivity analysis have a better correspondence with the expected system behavior.


\section{Conclusions}

Computer-based models for real-life processes \change{often}  consist of systems of numerous nonlinear equations, with deterministic, as well as stochastic variables. Increases in the level of detail or accuracy within these models often imply an explosion in the number of degrees of freedom, sometimes to the point where a high number of simulations becomes unfeasible even on modern computing hardware.

This paper explored the prospect of performing sensitivity analysis on the Delft3D-WAQ sediment transport model, aiming to identify the parameters that have the strongest effects on the variability of the model predictions. The complexity and non-linearity of the model, along with the engagement of a great number of parameters, led to the application of the Morris method, due to its versatility and computational efficiency.

An extension to Morris' classical method was proposed, allowing it to incorporate prior information about the dependence structure between model parameters into the sampling strategy. The extended method introduces copulas, which can accommodate a wide range of  dependence constraints and are generally applicable. The sensitivity analysis results correspond well with the expected behavior and dynamics of sediment transport in shallow waters. More specifically, the analysis revealed that the critical shear stress and the factor responsible for re-suspension from the sandy layer $S_2$ have the highest impact on the variance of the output. Consequently, and after expert assessment, the results of this study were used as a screening tool for subsequent model calibration, where the $5$ significant pairs of parameters were subjected to a simulated annealing algorithm, in order to determine the optimal values which give the best fit between the model output and the remote sensing data.

The results of the sensitivity analysis applied for a set of dependent parameters demonstrate the potential use of the extended Morris method in determining the key driving factors of a complex model. The method may be representative for similar studies of complex models worldwide and has been implemented in a generic approach. To that end, the Matlab code used to obtain the results presented in this paper is openly available \citep{code}.

\change{The dependence structure of parameters in the Delft3D-WAQ sediment transport model was very specific (complete positive / negative dependence between pairs of factors). This type of dependencies lead to exactly the same sensitivity behavior of these factors. However the method can accommodate other dependence structures, as well as complicated joint distributions of factors in the model. It would be of great importance to test the methodology presented in this paper for different types of models.}

\section{Acknowledgements}
The authors would like to acknowledge Deltares for their openness in providing access to the Delft 3D-WAQ sediment transport model and their approval and technical support for the simulations necessary to obtain the results presented in this study.

\section*{References}
\bibliography{references}

\end{document}